\documentclass[prb,twocolumn,nofootinbib,showpacs,superscriptaddress]{revtex4-1} %
\usepackage[latin1]{inputenc}
\usepackage{graphicx}
\usepackage{amssymb}
\usepackage{amsmath}
\usepackage{xspace}
\usepackage{dcolumn}
\usepackage{bm}
\usepackage{color}
\usepackage[extra]{tipa}

\newfont{\tensy}{cmsy10}

\newcommand{\ie}[0]{i.e.\@\xspace}
\newcommand{\eg}[0]{e.g.\@\xspace}
\newcommand{\etal}[0]{{\em et al.}\@\xspace}

\newcommand{\UP}[0]{\uparrow}
\newcommand{\DO}[0]{\downarrow}

\newcommand{\on}{\hat{n}}
\newcommand{\oS}{\hat{S}}

\newcommand{\oh}{\mbox{$\frac{1}{2}$}}

\newcommand{\si}[0]{\sigma}
\newcommand{\om}[0]{\omega}

\newcommand{\kF}{k_\text{F}}
\newcommand{\kB}{k_\text{B}}
\newcommand{\nag}{{\phantom{\dag}}}

\newcommand{\las}[0]{\langle}
\newcommand{\ras}[0]{\rangle}

\newcommand{\la}[0]{\left\las}
\newcommand{\ra}[0]{\right\ras}
\newcommand{\ket}[1]{\left|#1\ra}
\newcommand{\bra}[1]{\la#1\right|}

\newcommand{\en}[0]{\epsilon}

\begin{document}

\title{Interaction-range effects for fermions in one dimension}

\author{Martin Hohenadler}
\affiliation{%
\mbox{Institut f\"ur Theoretische Physik und Astrophysik, Universit\"at W\"urzburg,
97074 W\"urzburg, Germany}}

\author{Stefan Wessel}
\affiliation{%
Institut f\"ur Theoretische Festk\"orperphysik, JARA-FIT and JARA-HPC, RWTH Aachen University, 52056 Aachen, Germany}

\author{Maria Daghofer}
\affiliation{%
Institut f\"ur Theoretische Festk\"orperphysik, IFW Dresden, 01171 Dresden, Germany}

\author{Fakher F. Assaad}
\affiliation{%
\mbox{Institut f\"ur Theoretische Physik und Astrophysik, Universit\"at W\"urzburg,
97074 W\"urzburg, Germany}}

\begin{abstract}
  Experiments on quasi-one-dimensional systems such as quantum wires and
  metallic chains on surfaces suggest the existence of electron-electron
  interactions of substantial range and hence physics beyond the Hubbard
  model. We therefore investigate one-dimensional, quarter-filled chains with
  a Coulomb potential with variable screening length by quantum Monte Carlo
  methods and exact diagonalization. The Luttinger liquid interaction
  parameter $K_\rho$ decreases with increasing interaction strength and
  range. Experimentally observed values close to 1/4 require strong
  interactions and/or large screening lengths.  As predicted by bosonization,
  we find a metal-insulator transition at $K_\rho=1/4$.  Upon increasing the
  screening length, the charge and spin correlation functions reveal the
  crossover from dominant $2\kF$ spin correlations to dominant $4\kF$ charge
  correlations, and a strong enhancement of the charge velocity. In the
  metallic phase, the signatures of spin-charge separation in the
  single-particle spectrum, spinon and holon bands, remain robust even for
  rather long-ranged interactions. The charge-density-wave state exhibits
  backfolded shadow bands.
\end{abstract} 

\date{\today}


\maketitle

\section{Introduction}\label{sec:introduction}

The Hubbard model has served as a framework to study strongly correlated
electrons for almost five decades.\cite{Hu63} Its relative simplicity
compared to more realistic models is largely based on approximating the
electron-electron Coulomb interaction by an onsite repulsion $U$ between
electrons of opposite spin. The resulting Hamiltonian captures many aspects
of strong correlations, including the Mott transition at half
filling. Detailed knowledge about the model can be obtained by combining the
Bethe ansatz with the bosonization technique.\cite{Giamarchi_book_04}
However, experiments on quasi-one-dimensional (1D) systems such as quantum
wires,\cite{Me.Ma.09} carbon nanotubes,\cite{De.Bo.08} or self-organized atom
chains \cite{AuchainsNature} fall outside the range of validity of the
Hubbard model. This is evinced by the possibility of an insulating,
charge-ordered state at quarter filling, substantial $4\kF$ charge
correlations, or by a Luttinger liquid (LL) interaction parameter smaller
than 1/2. Within a quasi-1D description, these features imply
electron-electron interactions of finite range.

The case of one dimension is particularly interesting due to the breakdown of
Fermi liquid theory, the importance of collective excitations, and the
emergence of spin-charge separation. These phenomena can be understood in the
framework of
bosonization,\cite{Haldane.81,Emery-review,0034-4885-58-9-002,Giamarchi_book_04}
which provides a description in terms of a few nonuniversal parameters valid
asymptotically at long wavelengths and low energies. In particular, knowledge
of these parameters fully characterizes the correlation functions.

The 1D Hubbard model, describing a screened, onsite interaction, is a Mott
insulator for any $U>0$ at half filling. Away from half filling, umklapp
scattering is not allowed and the system remains metallic. The LL interaction
parameter takes on values $1/2\leq K_\rho \leq1$, leading to dominant spin
density wave correlations. A finite interaction range permits Mott or
charge-density-wave (CDW) transitions of the Kosterlitz-Thouless type at
other commensurate fillings $n$, for example at quarter filling in the
$U$-$V$ model with onsite ($U$) and nearest-neighbor ($V$)
repulsion.\cite{Giamarchi_book_04} In contrast to the Hubbard model, such
transitions occur at a finite critical $U$ determined by the condition
$K_\rho=n^2$. The effects of extended-range interactions depend on the
details. For example, the intuitive picture of long-range interactions
driving the system to strong coupling does not always apply: for spinless
fermions, the critical interaction for the metal-CDW transition is larger for
the $1/r$ potential than for a nearest-neighbor
repulsion;\cite{PhysRevB.61.13410} for spinfull fermions, a transition seems
to be absent for the unscreened potential up to very strong
interactions.\cite{PhysRevB.60.15654}

The $1/r$ Coulomb potential realized in, \eg, nanotubes and quantum wires,
represents the extreme limit of long-range interactions. The logarithmic
divergence of its Fourier transform gives rise to remarkable differences,
most notably the metallic Wigner crystal (WC) state with quasi-long-range
$4\kF$ charge correlations,\cite{PhysRevB.45.8454,PhysRevLett.71.1864} and
the existence of plasmon excitations. Strictly speaking, the divergence only
exists for infinite systems and in the absence of screening. Consequently,
the above phenomena are absent for any large but finite interaction range,
and the bare Coulomb potential can be regarded as a special point in
parameter space distinct from the LL liquid fixed point. The $1/r$ potential
has been studied
analytically\cite{PhysRev.168.418,PhysRevB.17.494,PhysRevLett.71.1864,PhysRevB.19.6119,PhysRevLett.72.2235,PhysRevLett.77.1358,PhysRevB.61.15530,Be.Go.00,PhysRevB.64.193307,PhysRevB.68.045112}
and
numerically.\cite{PhysRevB.60.15654,PhysRevB.61.13410,PhysRevB.75.125116,Da.No.Ho.08}

The typical experimental situation is most likely intermediate between the
Hubbard limit and the bare $1/r$ potential. Within bosonization, a finite
interaction range only leads to a renormalization of the LL
parameters.\cite{Giamarchi_book_04,0953-8984-1-42-018} However, in contrast
to the Hubbard model, there exist no analytical methods to calculate the LL
parameters exactly for nontrivial cases. Besides, the bosonization results
rely on a linear band dispersion, and are valid only at low energies and long
wavelengths, a limit which is nontrivial to achieve both in experiment and in
numerical simulations. On the other hand, exact numerical methods are valid
at all energies and distances and permit, \eg, the calculation of spectral
weights of excitations. They provide a quantitative connection to microscopic
model parameters, and can be used to study intermediate interaction
ranges. The 1D nature of the problem makes numerical methods particularly
powerful.

In this work we study the effect of the electron-electron interaction range
using exact, large-scale quantum Monte Carlo (QMC) simulations and exact
diagonalization. The model chosen here makes significant simplifications over
typical experimental situations, but we believe that our findings are rather
general. One of the key results is the LL interaction parameter $K_\rho$,
which allows us to estimate the interaction strength and range required to
reproduce the experimentally observed values.  We also study the evolution of
static and dynamical correlation functions as a function of the interaction
range. Importantly, we find that spin-charge separation in the
single-particle spectrum is robust against increasing the interaction
range. Our work extends previous investigations of spinfull and spinless
lattice
models,\cite{PhysRevB.17.494,PhysRevB.56.R1645,PhysRevB.61.13410,PhysRevB.68.045112,PhysRevB.75.125116,PhysRevB.69.195115}
and continuum simulations.\cite{PhysRevB.78.165303,PhysRevB.83.153303,PhysRevB.83.245114}

The paper is organized in the following way. In Sec.~\ref{sec:model} we
introduce the model and discuss related previous
work. Section~\ref{sec:methods} gives details of the numerical methods.  Our
results are discussed in Sec.~\ref{sec:results}. Sec.~\ref{sec:conclusions}
contains the conclusions. The appendix provides details about the application
of the continuous-time (CT)QMC method.

\section{Model}\label{sec:model}

We consider a 1D chain of length $L$ with Hamiltonian
\begin{equation}\label{eq:ham}
  \hat{H} 
  =
  \sum_{k} \epsilon(k) \on_{k} 
  + 
  \sum_{r=0}^{L/2-1} V(r) \sum_{i=1}^L \on_{i} \on_{i+r}\,.
\end{equation}
The kinetic term contains the usual 1D tight-binding band structure,
$\en(k)=-2t\cos k$. The electron density operator (summed over spin $\sigma$)
at wavevector $k$ (Wannier site $i$) is given by $\on_k$ ($\on_i$), with
$\on_{i\sigma}=c^\dag_{i\sigma}c^\nag_{i\sigma}$. We have set the lattice
constant, $\hbar$ and $\kB$ equal to one, and take $t$ as the unit of energy.

The interaction matrix element $V(r)$ is defined as 
\begin{equation}\label{eq:V}
V(r) = 
\begin{cases}
  V\,, & r=0\,,\\
  Ve^{-r/\xi}/2r\,, & r>0\,.
\end{cases}
\end{equation}
The screening length $\xi$ permits us to interpolate between the Hubbard
model ($\xi=0$, $U=2V$) and long-range Coulomb interaction [$\xi=\infty$,
$V(r)\sim 1/r$]. The choice~(\ref{eq:V}) appears more natural than gradually
adding more and more matrix elements for increasing distances. The condition
$r<L/2$ is due to the use of periodic boundary conditions. $V(r)$ as defined
by Eq.~(\ref{eq:V}) satisfies $V(r)\to0$ as $r\to\infty$ as well as the
convexity condition $V(r+1)+V(r-1)\geq 2V(r)$ for $r>1$. In the classical
limit (no hopping), this guarantees a $4\kF$ CDW ground
state.\cite{PhysRevB.17.494} If the second condition is not met, the
competition between $2\kF$ and $4\kF$ charge order can lead to enhanced
metallic behavior or even a CDW-metal transition with increasing interaction
range, as observed in quarter-filled extended Hubbard
models.\cite{PhysRevB.56.R1645,PhysRevB.72.033101} As we show below, our
choice of $V(r)$ excludes such phenomena. We have also compared
the choice of potential~(\ref{eq:V}) to an Ewald summation for the case
of Fig.~\ref{fig:akw-vs-xi-V6}, where the cutoff is expected to be most
relevant, but found only minor changes in the form of energy shifts.

The bosonization picture for the model~(\ref{eq:ham}), taking into account
the lattice, is as follows. At half filling, any $V>0$ produces a Mott
insulator. For commensurate densities $n$ away from half filling and an
interaction range greater than or equal to the average particle spacing
$1/n$, strong enough interactions cause a CDW transition at the critical
point $K_\rho=n^2$, beyond which umklapp scattering becomes a relevant
perturbation.\cite{Giamarchi_book_04} The CDW state is characterized by
long-range $4\kF$ charge order. In the LL phase, the dominant correlations
are $2\kF$ spin-density fluctuations for $K_\rho>1/3$, and $4\kF$ charge
correlations for $K_\rho<1/3$.  For the unscreened Coulomb potential with
divergent Fourier transform ($\xi=\infty$), we formally have $K_\rho=0$,
which would suggest an insulating ground state, in contrast to the continuum
prediction of a metallic quasi-WC made by Schulz.\cite{PhysRevLett.71.1864}

The existence of a metal-insulator transition at $K_\rho=n^2$ has been
verified numerically for the $U$-$V$ model and the $U$-$V_1$-$V_2$ model.  In
contrast, for lattice fermions with a long-range potential (more
specifically, the Pariser-Parr-Pople model), numerical
results\cite{PhysRevB.60.15654} suggest a metallic ground state with the
properties predicted in the absence of umklapp
scattering.\cite{PhysRevLett.71.1864} This rather surprising result, obtained
on large but finite systems, is attributed to the reduction of the umklapp
matrix element $g_3$ due to long-range interactions.\cite{PhysRevB.61.13410}
Within bosonization, there are subtle but important differences between
spinfull and spinless models (concerning umklapp scattering), and between odd
and even filling factors (\eg, $n=1/2$ and $n=1/3$ are not equivalent when
considering the Luther-Emery point).\cite{Giamarchi_book_04} These
differences seem to manifest themselves also in numerical studies of lattice
models. For example, whereas spinfull fermions interacting via a $1/r$
potential remain metallic even for large $V$,\cite{PhysRevB.60.15654} a
metal-insulator transition has been observed in the spinless
case,\cite{PhysRevB.56.R1645} with the critical interaction being larger than
for the extended Hubbard model.

For simplicity, we consider in the following only the case $\xi<\infty$, so
that no divergence in the Fourier transform $V(q)$ occurs. We further focus
on quarter filling $n=0.5$, and will see below that the model~(\ref{eq:ham})
is then either a LL (for $K_\rho>1/4$) or a CDW insulator (for $K_\rho<
1/4$).

For quarter filling, $n=0.5$, most of the physics of the model~(\ref{eq:ham})
(with $\xi<\infty$) can also be captured by simpler $U$-$V$ or
$U$-$V_1$-$V_2$ models provided the convexity condition is
satisfied.\cite{PhysRevB.17.494} In particular, these models realize the
non-Hubbard regime $K_\rho<1/2$, and a metal-insulator transition at
$K=1/4$. In the metallic phase, the LL conjecture implies that given the same
LL parameters, the extended Hubbard models and Eq.~(\ref{eq:ham}) produce
identical results, albeit with different microscopic parameters. However, in
connection with experiments, it is crucial to know how strong the dependence
of the LL parameters and hence the static and dynamical correlation functions
on the interaction range is. We will show below that in order to reach the
same value of $K_\rho$, the $U$-$V$ model requires much larger (and thus
rather unrealistic) interactions than a model with a larger interaction
range.

\section{Methods and observables}\label{sec:methods}

The majority of our results were obtained from simulations in the stochastic
series expansion (SSE) representation with directed loop
updates.\cite{SandvikSSE1,SySa02} The inclusion of the long-range interaction
terms in Eq.~(\ref{eq:V}) is straight forward. Due to the linear scaling of
computing time with the average expansion order, this method permits us to study
low temperatures and long chains (up to $L=140$ here) even in the
strong-coupling regime. We also show results obtained with the
continuous-time QMC method.\cite{Ru.Sa.Li.05,assaad:035116} The latter is
restricted to weak and intermediate interactions due to a less favorable
scaling of computer time with temperature and system size, and additional
numerical difficulties (see the Appendix). Both QMC methods are exact.

The single-particle spectral function is of particular interest in relation
to photoemission results. Since the calculation of the single-particle Green's
function in SSE is hampered by a minus-sign problem (for periodic
boundaries), we instead present results from exact diagonalization on
clusters with $L=20$.

We consider the static charge ($\rho$) and spin ($\sigma$) structure factors
\begin{align}\nonumber\label{eq:static}
  S_\rho(q)
  &=
  \sum_{r} e^{iq r} 
  \left(
  \las \on_r \on_0\ras
  -\las n_r\ras\las n_0\ras
  \right)
  \,,
  \\
  S_\sigma(q)  
  &=
  \sum_{r} e^{iq r} 
  \las \oS^z_r \oS^z_0\ras
  \,,
\end{align}
where $\oS^z_j=\oh(\on_{j\UP}-\on_{j\DO})$, and the dynamical charge and spin structure factors
\begin{align}\label{eq:dynamic}
  S_{\rho}(q,\om)
  &=
  \frac{1}{Z}\sum_{ij}
  e^{-\beta E_j} {|\bra{i} {\hat{\rho}}_q \ket{j}|}^2 \delta(E_i-E_j-\om)
  \,,
  \\
  S_{\sigma}(q,\om)
  &=
  \frac{1}{Z}\sum_{ij}
  e^{-\beta E_j} {|\bra{i} \oS^z_q \ket{j}|}^2 \delta(E_i-E_j-\om)\,,
\end{align}
where $\hat{\rho}_q = \sum_r e^{iqr} (\on_{r} - n) /\sqrt{L}$,
and $\ket{i}$ and $\ket{j}$ are eigenstates with energies $E_i$ and $E_j$.
These dynamical correlation functions can be calculated in the SSE
representation at fixed particle density and for periodic boundaries without
a sign problem. For the analytical continuation we have used the maximum
entropy method.\cite{Beach04a}

The $T=0$ single-particle spectral function reads as
\begin{align}\label{eq:akw}
  A(k,\om)
  &=
  A^+(k,\om) + A^-(k,\om)\,,\\\nonumber
  A^+(k,\om) &= 
  \sum_{n}
  {|\langle{\psi^{(N_\text{e}+1)}_{n,k}}| c^\dag_{k-q}
    |{\psi^{(N_\text{e})}_{0,q}}\rangle|}^2 \\\nonumber
  &\quad\quad\quad\quad\times\delta\left[\om -
    \left(E^{(N_\text{e}+1)}_{n,k}-E^{(N_\text{e})}_{0,q}\right)\right]
  \,,
  \\\nonumber
   A^-(k,\om) &= 
  \sum_{n}
  {|\langle{\psi^{(N_\text{e}-1)}_{n,k}}| c^\nag_{k-q}
    |{\psi^{(N_\text{e})}_{0,q}}\rangle|}^2 \\\nonumber
  &\quad\quad\quad\quad\times\delta\left[\om +
    \left(E^{(N_\text{e}-1)}_{n,k}-E^{(N_\text{e})}_{0,q}\right)\right]
  \,,
\end{align}
where $A^-$ ($A^+$) is related to photoemission (inverse photoemission), and
$|\psi^{(N_\text{e})}_{0,k}\rangle$ denotes the groundstate for the sector
with $N_\text{e}$ electrons and total momentum $k$; the corresponding
energy is $E^{(N_\text{e})}_{0,k}$. In order to measure energies relative to the
Fermi energy, we show $A(k,\om-\mu)$ with $\mu=[E_0^{(N_\text{e}+1)}-E_0^{(N_\text{e}-1)}]/2$.

\section{Results}\label{sec:results}

\begin{figure}[t]
  \includegraphics[width=0.45\textwidth]{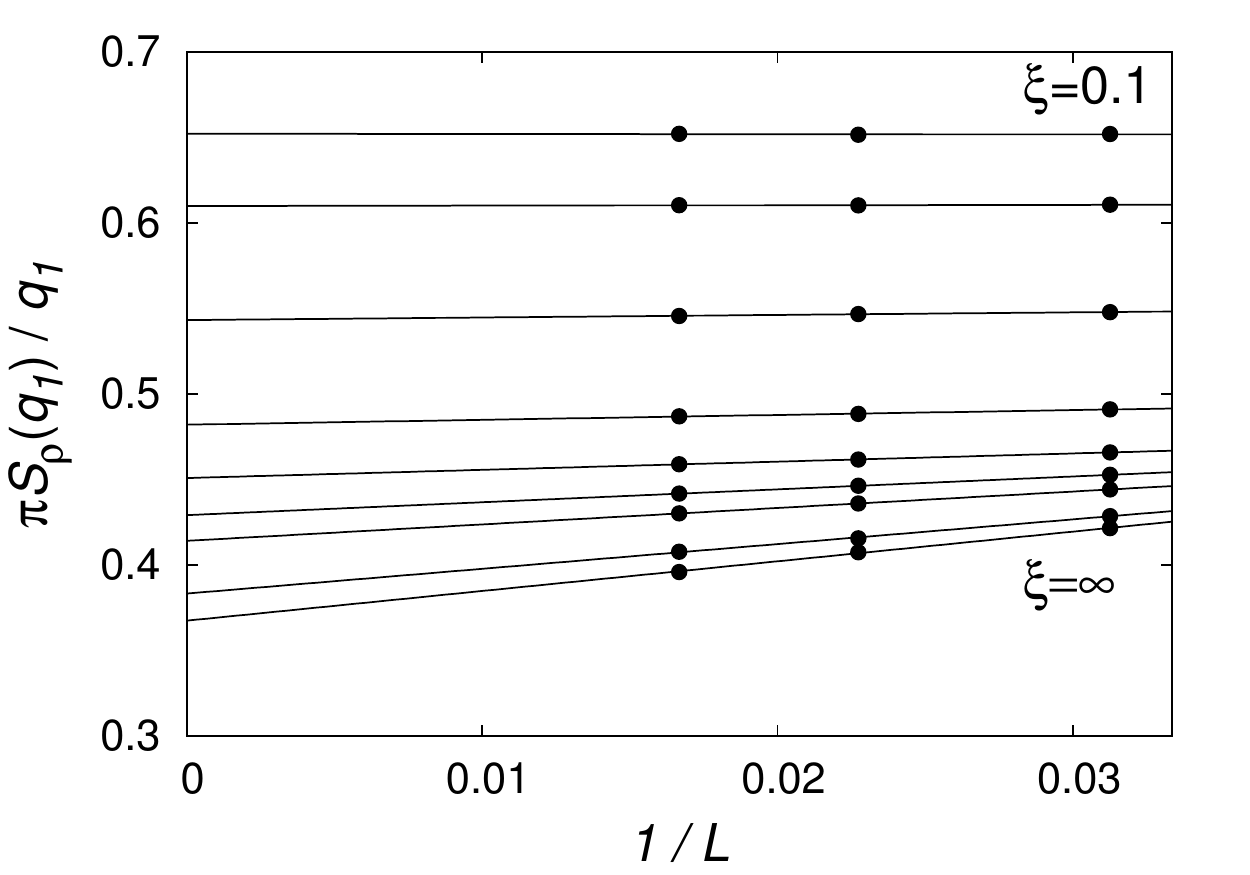}
  \caption{\label{fig:krho2} Finite-size scaling of the rescaled density
    structure factor $\pi S_\rho(q_1)/q_1$, with $q_1=2\pi/L$,
    for $V/t=3$, $n=0.5$ and $\xi=0.1$, 0.5, 1, 2, 3, 4,
    5, 10, 20 (top to bottom). Lines are linear fits, and the extrapolated value in the thermodynamic limit
    $L\to\infty$ defines the Luttinger liquid parameter $K_\rho$.
    The extrapolation has
    been carried out for all data points shown in Fig.~\ref{fig:krho}. The
    system sizes were $L=32,44,60$ for $n=0.5$ and $L=60,100,140$ for
    $n=0.1$. The temperature was $\beta t=2L$ for $V/t=1$ and $\beta t=L$ for $V/t=3,6,9$.}
\end{figure}

\begin{figure}[t]
  \includegraphics[width=0.45\textwidth]{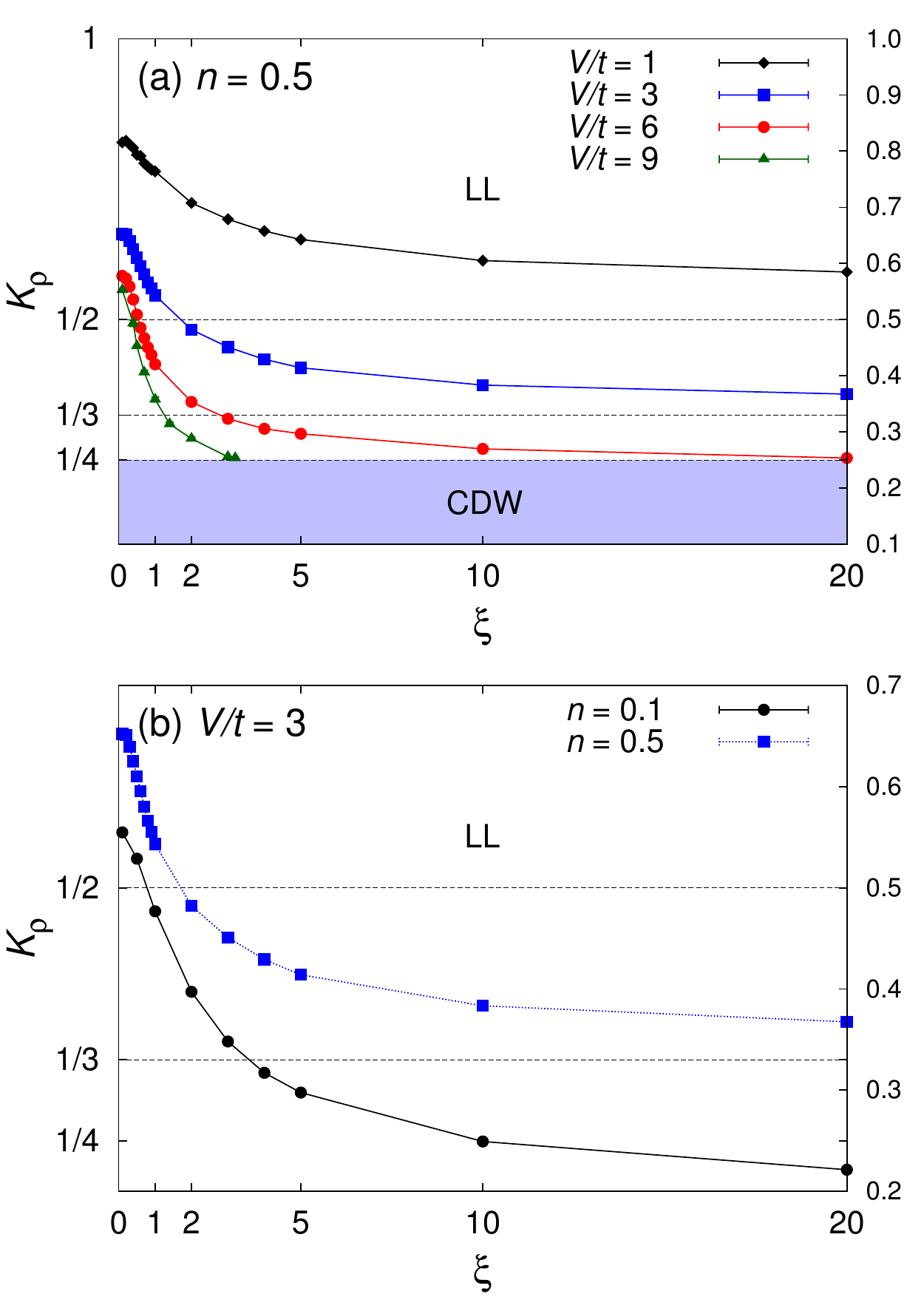}
  \caption{\label{fig:krho} (Color online) Luttinger liquid parameter
    $K_\rho$ as a function of screening length $\xi$. Points represent values
    obtained from a finite-size scaling (see Fig.~\ref{fig:krho2}). Lines are
    guides to the eye. (a) Results at quarter filling $n=0.5$. (b) Fixed
    $V/t=3$ and different fillings $n$. The phases are a Luttinger
    liquid (LL) for $K_\rho> n^2$ and a charge-density-wave insulator (CDW)
    for $K_\rho<n^2$.
  }
\end{figure}

Since we used three different methods, let us state here that the results of
Figs.~\ref{fig:krho2}--\ref{fig:lrorder}, \ref{fig:dynamics-vs-xi-V6}
and~\ref{fig:dynamics-vs-xi-V9} were obtained using the SSE representation,
Fig.~\ref{fig:dynamics-vs-xi-V1} with the CTQMC method, and
Figs.~\ref{fig:akw-vs-xi-V1}--\ref{fig:akw-vs-xi-V9} by exact
diagonalization. Except for Fig.~\ref{fig:krho}(b), results are for quarter
filling $n=0.5$.

\subsection{Luttinger liquid interaction parameter}

In the metallic regime of the model~(\ref{eq:ham}), the knowledge of the LL
interaction parameter $K_\rho$ together with the bosonization results for the
correlation functions provides a complete description of the low-energy,
long-wavelength physics. The crossover between the Hubbard and long-range
cases as a function of $\xi$, and the quantitative relation between
microscopic parameters and LL parameters, can be studied exactly by means of
numerical methods. The LL parameter has previously been calculated, for
example, for spinless fermions with a $1/r$
potential,\cite{PhysRevB.61.13410} for the $U$-$V$
model,\cite{0295-5075-70-4-492} and for the $U$-$V_1$-$V_2$
model.\cite{PhysRevB.72.033101}

We extract $K_\rho$ from SSE QMC results for the charge
structure factor using the relation
\begin{equation}
K_{\rho} = \lim_{L\to\infty} \frac{\pi}{q_1} S_{\rho}(q_1)\,,
\end{equation}
where $q_1=2\pi/L$ is the smallest, nonzero wavevector for a given system
size, and the static structure factor is defined in Eq.~(\ref{eq:static}).
For each $V$ and $\xi$, we have performed a finite-size scaling to obtain $K_\rho$.
The extrapolation is shown for selected values of $\xi$ in the case $V/t=3$,
$n=0.5$ in Fig.~\ref{fig:krho2}. We find that for large enough
system sizes, the finite-size dependence is dominated by the lowest order
$1/L$, and have therefore used a linear fit for the extrapolation.

Figure~\ref{fig:krho}(a) shows the dependence of $K_\rho$ on $V/t$ and $\xi$
at quarter filling $n=0.5$. The $V/t=1$ results fall into the Hubbard regime
$K_\rho\geq1/2$ for all values of $\xi$ shown. For a stronger interaction
$V/t=3$, $K_\rho$ becomes smaller than 1/2 for $\xi\approx2$, but remains
larger than 1/3, thereby implying dominant $2\kF$ correlations [see
Eq.~(\ref{eq:correl_Hubbard}) and discussion below].  At $V/t=6$, the values
of $K_\rho$ span the Hubbard, non-Hubbard and dominant $4\kF$ (\ie,
$K_\rho<1/3$) regimes. For the largest $\xi=20$, the LL parameter takes on
almost exactly the critical value $K_\rho=1/4$ of the LL-CDW transition. The
numerical results therefore suggest that the experimentally observed values
of $K_\rho\approx0.25$ require surprisingly large values of $V/t$ and
$\xi$. Finally, for $V/t=9$, the system undergoes the metal-insulator
transition for $\xi\approx3.5$. Independent of $V$, we expect $K_\rho\to0$
for $\xi\to\infty$ in the thermodynamic limit, corresponding to the
quasi-WC. A theoretical prediction, $K_\rho\sim \ln^{-1/2}\xi$, was made by
Schulz,\cite{PhysRevLett.71.1864} and the numerical results for the charge
structure factor by Fano \etal\cite{PhysRevB.60.15654} are consistent with
$K_\rho=0$.

Figure~\ref{fig:krho}(a) reveals that $K_\rho$ decreases with increasing
$\xi$, thereby bringing the system closer to the insulating phase. In
previous work on extended Hubbard models, it was found that adding
interactions at distances beyond the interparticle spacing $1/n$ can increase
$K_\rho$ and hence enhance the metallic character of the
system.\cite{PhysRevB.56.R1645,PhysRevB.61.13410} Similarly, in the
$U$-$V_1$-$V_2$ model with $U$ fixed, varying the relative strength of $V_1$
and $V_2$ leads to a competition between $2\kF$ and $4\kF$ charge
fluctuations.\cite{PhysRevB.56.R1645,PhysRevB.72.033101} As a result,
$K_\rho$ takes on a maximum for $V_2=V_1/2$, where the metallic state is most
stable, and it is not clear if the $U$-$V_1$-$V_2$ becomes insulating at
finite values of $V_1$ and $V_2$.\cite{PhysRevB.72.033101,PhysRevB.69.195115}
The condition $V(2)=V(1)/2$ is also realized for the unscreened Coulomb
potential, and numerical results suggest that the system remains metallic up
to very strong interactions even in the presence of a
lattice.\cite{PhysRevB.56.R1645,PhysRevB.60.15654} The experimentally
motivated form~(\ref{eq:V}), fulfilling the monotonicity and the convexity
condition,\cite{PhysRevB.17.494} favors a $4\kF$ CDW state in the limit
$V/t\to\infty$.\cite{PhysRevB.17.494} Similar to previous results for
spinless fermions with a $1/r$ potential,\cite{PhysRevB.61.13410} $K_\rho$ in
Fig.~\ref{fig:krho} decreases with increasing $V/t$.

A common feature of the curves in Fig.~\ref{fig:krho}(a) is a pronounced
decrease at small values of $\xi$, followed by a much slower decrease for
larger $\xi$. The numerical results indicate that the change in behavior
occurs when the interaction range $\xi$ equals the average particle spacing
$1/n=2$. To verify this hypothesis, we compare in Fig.~\ref{fig:krho}(b) the
$\xi$ dependence of $K_\rho$ for two different densities $n=0.5$ and $n=0.1$
at $V/t=3$. The curve for $n=0.1$ indeed exhibits a significant $\xi$
dependence up to much larger $\xi$. The results for $n=0.1$ further reveal
that for a given $V/t$, a smaller density requires a significantly larger
interaction strength and/or range to reach the critical $K_\rho=n^2$ for the
metal-insulator transition, see also Ref.~\onlinecite{PhysRevB.68.045112}.

\subsection{Charge and spin correlation functions}

For a model with $SU(2)$ spin symmetry such as Hamiltonian~(\ref{eq:ham}),
bosonization predicts the decay of charge and spin correlation
functions to be determined solely by the parameter $K_\rho$
(since $K_\si=1$),\cite{PhysRevLett.64.2831} 
\begin{align}\label{eq:correl_Hubbard}\nonumber
  \las n_x n_0 \ras 
  &=-
  \frac{K_\rho}{(\pi x)^2}+ \frac{A_1}{x^{1+K_\rho}}
  \cos(2\kF x)  
  +
  \frac{A_2}{x^{4K_\rho}}\cos(4\kF x)\,,
   \\
   \las \oS^z_x \oS^z_0 \ras 
   &=-
  \frac{1}{(2\pi x)^2} + \frac{B_1}{x^{1+K_\rho}}\cos(2\kF x) 
  \,.
\end{align}
The $1/x^2$ dependence of the leading term in both channels is familiar from
Fermi liquid theory. The $2\kF$ and $4\kF$ charge correlations decay to
leading order as ${x^{-1-K_\rho}}$ and ${x^{-4K_\rho}}$, respectively. For the
Hubbard model, $K_\rho\geq 1/2$ and $2\kF$ correlations dominate; taking into
account logarithmic corrections not included in
Eq.~(\ref{eq:correl_Hubbard}), the dominant correlations in this regime are
$2\kF$ spin correlations.\cite{Giamarchi_book_04} For models with a nonzero
interaction range, the $4\kF$ density oscillations can become dominant for
$K_\rho<1/3$.  For the Hubbard model, subdominant $4\kF$ oscillations have
been observed in systems with open boundary
conditions.\cite{PhysRevB.79.195114}

\begin{figure}[t]
  \includegraphics[width=0.45\textwidth]{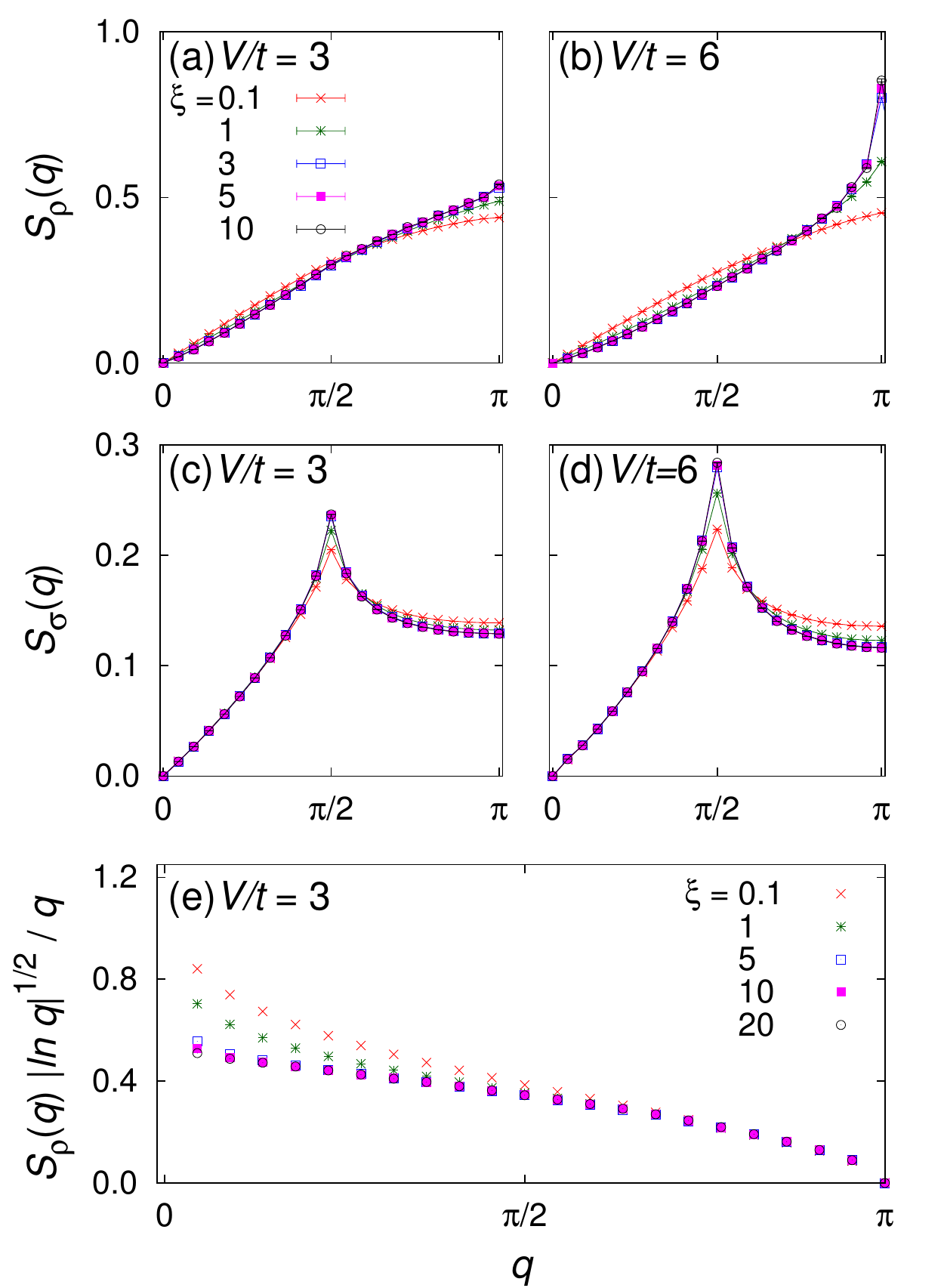}
  \caption{\label{fig:sf} (Color online) Charge [(a), (b)] and spin [(c),
    (d)] structure factors $S_{\rho/\sigma}(q)$ for different values of the
    screening length $\xi$, and $\beta t=L=44$. The key in (a) applies to
    (a)--(d). (e) Rescaled charge structure factor for $V/t=3$ ($\beta
    t=L=60$), revealing the asymptotic approach to the small-$q$ behavior of
    the WC, $S_\rho(q)\sim q/|\ln q|^{1/2}$. Here and in all subsequent
    figures, results are for quarter filling $n=0.5$.}
\end{figure}

\begin{figure}
   \includegraphics[width=0.45\textwidth]{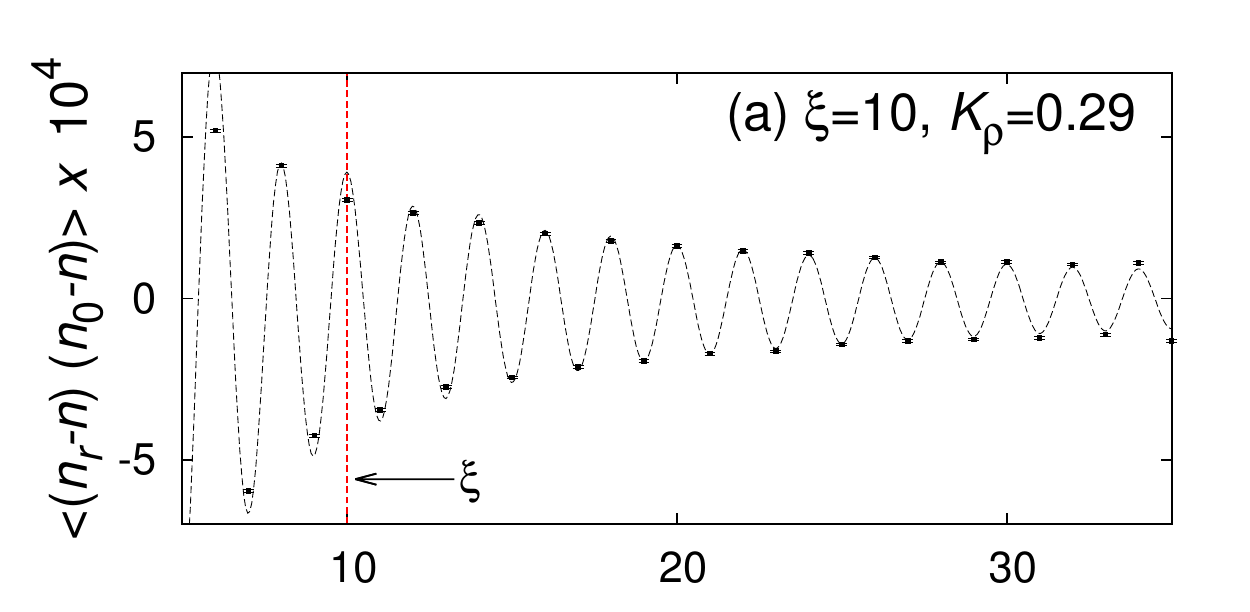}
   \includegraphics[width=0.45\textwidth]{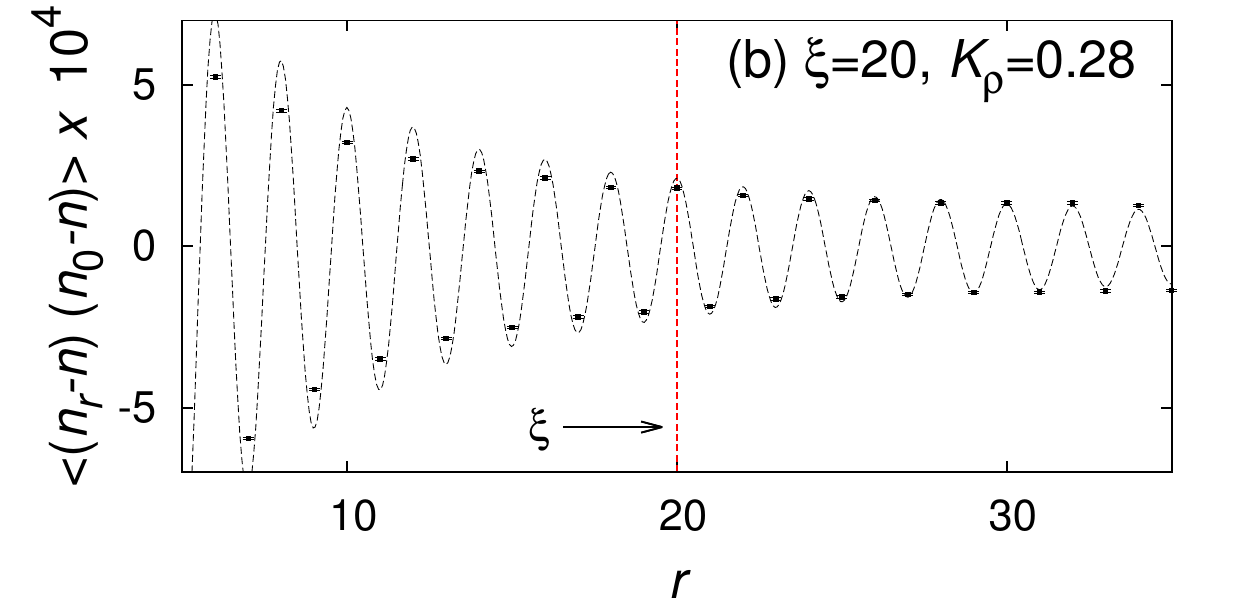}
   \caption{\label{fig:dens-realspace} 
     (Color online)
     Density-density correlations in real space (symbols). Here $V/t=5$,
     $\beta t=L=84$, $n=0.5$ and (a) $\xi=10$, (b) $\xi=20$. Lines are fits to the LL
     result for $\las n_x n_0\ras$ with fitting parameters $A_1$, $A_2$ [see
     Eq.~(\ref{eq:correl_lr})] and $K_{\rho}$ determined from a (linear) finite-size
     extrapolation based on $L=44,84$. The fitting interval was (a)
     $[15:35]$, (b) $[25:45]$. 
   }
\end{figure}

In the opposite limit of a $1/r$ Coulomb potential ($\xi=\infty$) with
divergent Fourier transform $V(q)\sim \ln(1/q)$, Schulz\cite{PhysRevLett.71.1864}
obtained
\begin{align}\label{eq:correl_lr}\nonumber
  \las \on_x \on_0 \ras 
  &=
  \frac{C_1}{x}
  e^{-c_2\sqrt{\ln x}}\cos(2\kF x)
  \\\nonumber
  &\quad\quad\quad\quad
  + 
  C_2 e^{-4c_2\sqrt{\ln x}}\cos(4\kF x)\,,
  \\
  \las \oS^z_x \oS^z_0 \ras 
  &=
  \frac{D_1}{x}
  e^{-c_2\sqrt{\ln x}} \cos(2\kF x)
  \,.
\end{align}
Apart from the absence of the $1/x^2$ Fermi liquid contribution, the most
notable difference is that charge correlations are dominated by an unusually
slow decay of the $4\kF$ component (slower than any power law).  These
quasi-long-range $4\kF$ charge oscillations led to the notion of a
fluctuating WC, where the wavelength $\lambda=2\pi/4\kF=1/n$ is the average
distance between fermions. In contrast, the spin sector retains a power-law
decay.  These continuum results are consistent with numerical
work.\cite{PhysRevB.78.165303,PhysRevB.60.15654}

As emphasized before, the WC results~(\ref{eq:correl_lr}) rely on the
divergence of the Fourier transform of the potential $V(r)$. Such a
divergence only occurs in the thermodynamic limit, and for $\xi=\infty$. If
either of these conditions is not met, the LL forms~(\ref{eq:correl_Hubbard})
can be recovered in the long-wavelength limit.  Here we only consider large
but finite values of $\xi$, for which a metal-insulator transition occurs at
$K_\rho=n^2=1/4$. The CDW state exhibits long-range $4\kF$ charge order. The
closest analog of the metallic quasi-WC state in our case is therefore the
metallic regime $1/3>K_\rho>1/4$ with dominant (power-law) $4\kF$
correlations. As shown in Fig.~\ref{fig:krho}, $K_\rho<1/3$ is realized for
$V/t=6$ and large $\xi$, and we explore the similarities to the WC below.

Figure~\ref{fig:sf} shows the charge and spin structure factors as defined in
Eq.~(\ref{eq:static}). At $V/t=3$ and with increasing $\xi$, we see a slight
increase of the $4\kF=\pi$ charge correlations, see
Fig.~\ref{fig:sf}(a). This effect becomes more noticeable for a stronger
repulsion $V/t=6$, as shown in Fig.~\ref{fig:sf}(b).  The inherent length
scale $1/n$ again appears in Fig.~\ref{fig:sf}, with the results saturating
on the scale of the plots for $\xi\gtrsim 2$. The spin structure factor
[Figs.~\ref{fig:sf}(c) and (d)] reveals an enhancement of $2\kF=\pi/2$
antiferromagnetic correlations with increasing $\xi$, which according to
Eq.~(\ref{eq:correl_Hubbard}) can be related to the reduction of
$K_\rho$. This enhancement is again more pronounced for $V/t=6$ than for
$V/t=3$.

Let us now turn to the long-wavelength behavior. For a LL we have
$S_\rho(q)\sim q K_\rho$, whereas for the WC $S_\rho(q)\sim q |\ln q|^{-1/2}$
(see Ref.~\onlinecite{Giamarchi_book_04}). Following
Ref.~\onlinecite{PhysRevB.60.15654}, we plot in Fig.~\ref{fig:sf}(e)
$S_\rho(q) |\ln q|^{1/2}/q$. This quantity shows a logarithmic divergence at
$q=0$ as long as $S_\rho(q)\sim q$ and tends to a constant as $q\to0$ for
$\xi=\infty$.\cite{PhysRevB.60.15654} Our numerical results show that a
divergence occurs throughout the metallic phase, and that the approach to the
WC result is rather slow.  In particular, given the finite values of $\xi$,
the LL nature of the system reemerges eventually in the limit $q\to0$,
although the system sizes required to see this effect become larger and
larger. A nonlinear (at long wavelengths) density structure factor
corresponding to $K_\rho=0$ has been observed for the $1/r$
potential.\cite{PhysRevB.60.15654} In contrast, for finite $\xi$,
Fig.~\ref{fig:sf} shows that the linear behavior of $S_\rho(q)$ is
preserved. The long-wavelength spin structure factor is not affected by the
interactions [Fig.~\ref{fig:sf}(c) and (d)]; the slope in the limit $q\to0$
remains fixed, as required by $K_\sigma=1$
[cf. Eq.~(\ref{eq:correl_Hubbard})].

\begin{figure}
  \includegraphics[width=0.45\textwidth]{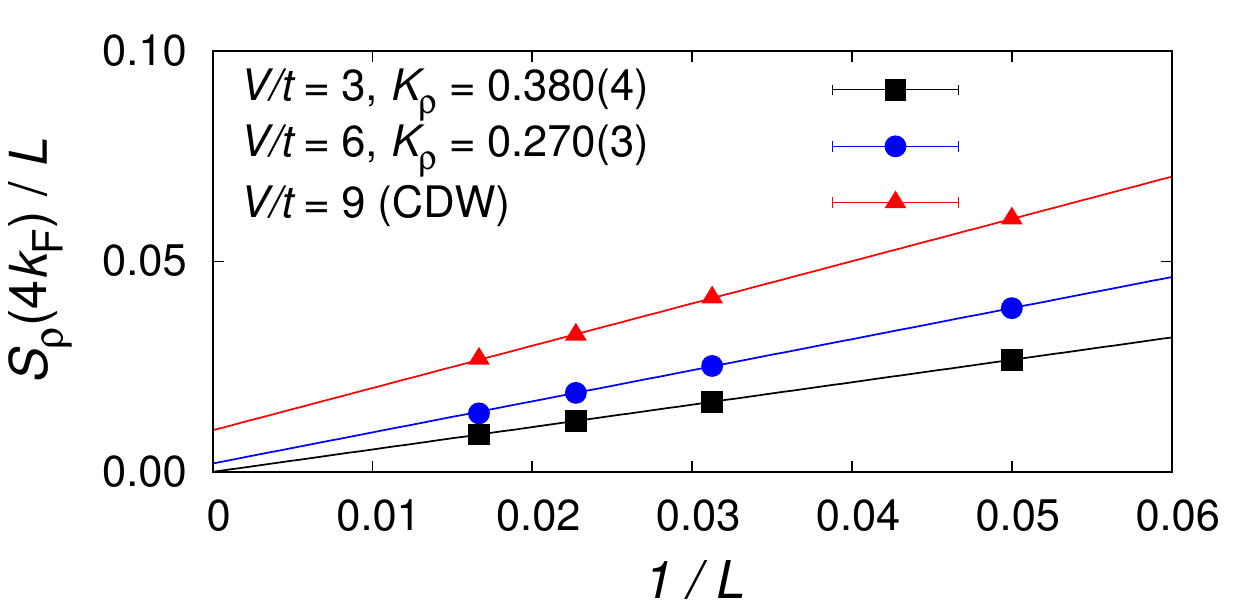}
  \caption{\label{fig:lrorder} 
    (Color online)
     Finite-size scaling of the amplitude of $4\kF$ charge
     correlations at fixed $\xi=10$. The results reveal the absence of
     long-range order for $V/t=3$, and long-range $4\kF$
     charge correlations for $V/t=9$. The case $V/t=6$ is on the metallic
     side but very close to the critical point, and we find a small but finite extrapolated
     value. The lines are linear fits. Here $\beta t=L$ and $n=0.5$.}
\end{figure}

Schulz\cite{PhysRevLett.71.1864} suggested that for a finite $\xi$, one
should be able to observe WC-like correlations at distances $x<\xi$ and
LL-like correlations at $x>\xi$. Although the bosonization results are only
valid for large distances, this prediction can in principle be tested
numerically. Figure~\ref{fig:dens-realspace} shows the density-density
correlation function in real space. We have chosen $V/t=5$, and $\xi=10$
or $\xi=20$. This choice was made for the following reasons.
First, deviations from the LL form given by Eq.~(\ref{eq:correl_Hubbard}) are
most visible in the regime where $4\kF$ oscillations dominate, that is for
$K_\rho<1/3$. However, for the bosonization results to apply, it is important
to avoid the insulating state expected for $K_\rho<1/4$. Close to
$K_\rho=1/4$, previous work on the extended ($U$-$V$) Hubbard model has shown
the importance of logarithmic corrections.\cite{PhysRevB.81.085116} For the
parameters chosen, we have $K_\rho\approx 0.29$ for $\xi=10$ and
$K_\rho\approx0.28$ for $\xi=20$. The results in
Fig.~\ref{fig:dens-realspace} show dominant $4\kF$ correlations but no
long-range order, as expected in the LL regime.

\begin{figure}
   \includegraphics[width=0.45\textwidth]{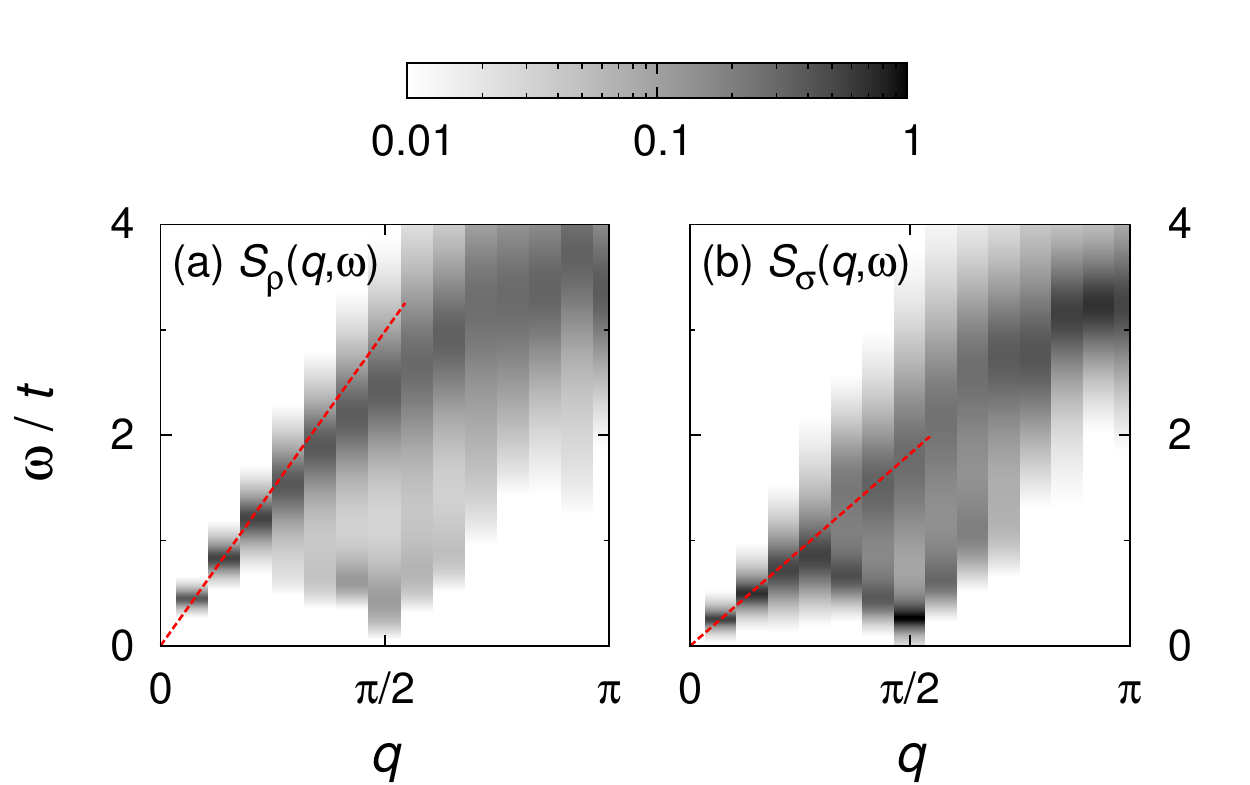}
   \caption{\label{fig:dynamics-vs-xi-V1} 
     (Color online)
     (a) Dynamical charge  and (b) spin structure factor  for $V/t=1$
     and $\xi=10$. Results were obtained
     with the projective CTQMC method\cite{assaad:035116} using $L=28$, $\theta t=15$ and
     $n=0.5$. Dashed lines indicate the  velocity of long-wavelength charge and spin excitations.
   }
\end{figure}

Based on the idea that the LL form for $\las n_x n_0\ras$ should hold at
distances larger than $\xi$, we fit the numerical data to
Eq.~(\ref{eq:correl_Hubbard}) using two fitting parameters (the $2\kF$ and
$4\kF$ amplitudes) as well as the above values of $K_\rho$. The fitting
intervals are chosen as $[\xi+5,35]$ and we used $\beta
t=L=84$. Figure~\ref{fig:dens-realspace}(a) shows that we indeed have good
agreement between the fit and the QMC data at large distances. However, for
$r\lesssim\xi=10$, significant deviations become visible. To discriminate
between short-distance effects coming from the continuum approximation
underlying Eq.~(\ref{eq:correl_Hubbard}) and genuine deviations from LL
theory we consider $\xi=20$ in Fig.~\ref{fig:dens-realspace}(b). Again there
is reasonable agreement at large distances, but clear differences at
$r\lesssim\xi=20$. Hence, keeping in mind the difficulties mentioned above,
our results are consistent with the picture proposed by
Schulz.\cite{PhysRevLett.71.1864}

As can be seen from Fig.~\ref{fig:krho}, the insulating CDW phase can be
reached for $\xi\gtrsim 3.5$ and $V/t=9$. The CDW state is characterized by
long-range $4\kF$ charge order at $T=0$, as formally reflected by
Eq.~(\ref{eq:correl_Hubbard}) for $K_\rho=0$, and may be regarded as a WC
pinned to the lattice. Figure~\ref{fig:lrorder} shows the amplitude of $4\kF$
charge correlations divided by system size, \ie, $S_\rho(4\kF)/L$.  At fixed
$\xi=10$, we find that this quantity extrapolates to zero in the
thermodynamic limit in the LL phase [$K_\rho=0.383(1)$, $V/t=3$], and to a
finite value in the CDW state ($V/t=9$). Near the phase boundary, the
Kosterlitz-Thouless nature of the transition makes numerical studies
difficult and we see that, assuming a linear scaling, $S_\rho(4\kF)/L$
extrapolates to a finite but very small value despite
$K_\rho=0.270(3)>1/4$. For the unscreened Coulomb potential, $S_\rho(4\kF)/L$
increases logarithmically with system size, and there is no long-range
order.\cite{PhysRevB.60.15654}

\begin{figure}
   \includegraphics[width=0.45\textwidth]{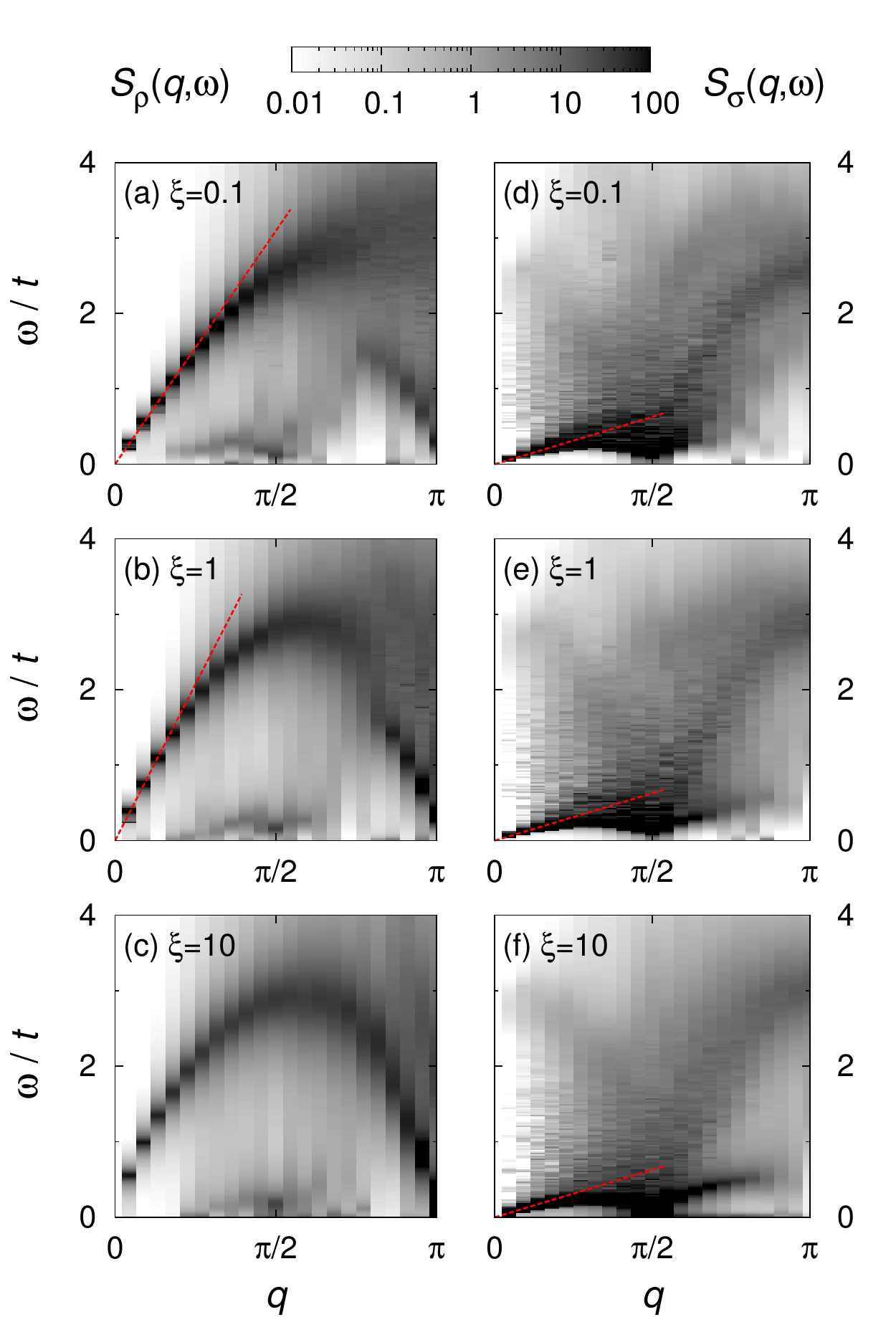}
   \caption{\label{fig:dynamics-vs-xi-V6} (Color online)
     Dynamical charge [(a)--(c)] and spin [(d)--(f)] structure factor
     from simulations in the SSE representation for $V/t=6$ and different screening lengths
     $\xi$. Results are for $n=0.5$, $L=44$ and $\beta t=2L$. The dashed lines
     indicate the velocity at long wavelengths.
   }
\end{figure}

\subsection{Dynamical charge and spin correlations}

We now discuss the dynamical spin and charge correlation functions, defined
in Eq.~(\ref{eq:dynamic}), as obtained from QMC simulations. We begin with a
rather weak interaction $V/t=1$ and a large screening length $\xi=10$.  CTQMC
results for these parameters which, according to Fig.~\ref{fig:krho}, fall
into the Hubbard regime, are presented in
Fig.~\ref{fig:dynamics-vs-xi-V1}. Despite the long-range interaction, the
spectra closely resemble previous results for the Hubbard model, see, \eg,
Ref.~\onlinecite{PhysRevB.73.165119}. In particular, the particle-hole
continuum is clearly visible in both the charge and the spin channels. As a
result of interactions, the velocities of long-wavelength charge and spin
excitations differ by about a factor of 2.

To investigate larger values of $V/t$, we use the SSE representation. The
latter can also be used for the parameters of Fig.~\ref{fig:lrorder}, but we
chose the CTQMC method to demonstrate its applicability to models with
long-range interactions. Taking $V/t=6$, we can explore the whole metallic
regime of the model~(\ref{eq:ham}) by varying the screening length
$\xi$. Results are shown in Fig.~\ref{fig:dynamics-vs-xi-V6}.

\begin{figure}
   \includegraphics[width=0.45\textwidth]{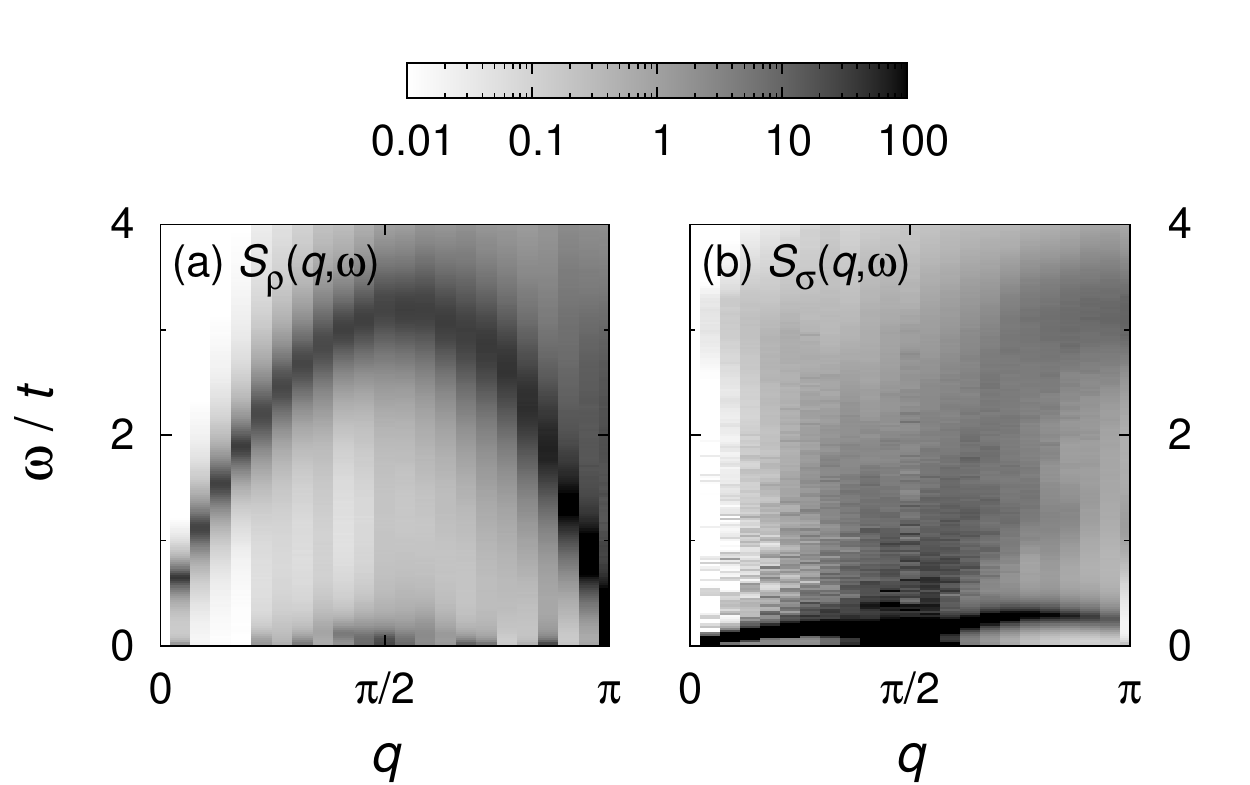}
   \caption{\label{fig:dynamics-vs-xi-V9} 
     (Color online)
     (a) Dynamical charge and (b) spin structure factor for $V/t=9$
     and $\xi=10$, corresponding to the insulating CDW phase. Results were
     obtained in the SSE representation using $n=0.5$,
     $L=44$ and $\beta t=2L$.     
   }
\end{figure}

We first discuss the charge sector. For $\xi=0.1$, corresponding to the
strong-coupling regime of the Hubbard model [$U=2V(0)=12t$], the results in
Fig.~\ref{fig:dynamics-vs-xi-V6}(a) look qualitatively similar to
Fig.~\ref{fig:dynamics-vs-xi-V1}(a). However, the distribution of spectral
weight over the particle-hole continuum is much more inhomogeneous, with
pronounced excitation features along the edges. The charge velocity $v_\rho$
is only slightly smaller than in Fig.~\ref{fig:dynamics-vs-xi-V1}(a).  Upon
increasing $\xi$, we observe a substantial increase of $v_\rho$, as indicated
by the dashed lines; between $\xi=0.1$ and $\xi=1$, $v_\rho$ increases from
$1.97(2)t$ to $2.64(2)t$.  A small charge gap of order $0.1t$, which
extrapolates to zero for $L\to\infty$ in the LL phase, is visible in
Fig.~\ref{fig:dynamics-vs-xi-V6}(c), but we can estimate the velocity as
$v_\rho>3.5t$. The increase of $v_\rho$ reflects the fact that the extended
interaction promotes $4\kF$ charge order, and thereby increases the stiffness
of the charges with respect to long-wavelength excitations.  This gap is a
finite-size effect caused by the close proximity of the CDW transition.  The
onset of $4\kF$ fluctuations is also reflected in an incomplete but well
visible softening of the excitations at $q=4\kF$. We will see below that this
feature develops into a Bragg peak in the CDW state. A plasmon excitation,
one of the hallmark features of the $1/r$ Coulomb potential, is not expected
for finite values of $\xi$, and would in general be very difficult to
distinguish from a linear mode in numerical simulations.

In contrast to the charge sector, the effect of $\xi$ on the spin dynamics is
very small. In accordance with LL theory, the velocity $v_\sigma$ of
long-wavelength spin excitations remains virtually unchanged upon increasing
$\xi$ from 0.1 to 10 [Fig.~\ref{fig:dynamics-vs-xi-V6}(d) and (f)]. However,
$v_\sigma$ is strongly renormalized in going from $V/t=1$
[Fig.~\ref{fig:dynamics-vs-xi-V1}(b)] to $V/t=6$
[Fig.~\ref{fig:dynamics-vs-xi-V6}(d)].  At fixed $V/t$, the screening length
hence provides a natural way of changing the ratio of charge and spin energy
scales, and opens a route to explore the spin-incoherent
LL.\cite{RevModPhys.79.801}

Figure~\ref{fig:dynamics-vs-xi-V9} shows results for the charge and spin
dynamics in the CDW phase, for $V/t=9$ and $\xi=10$. As demonstrated in
Fig.~\ref{fig:lrorder}, for these parameters, the system is in a CDW state
with long-range $4\kF$ order. In addition to a  charge gap at $q=0$, the
charge structure factor has become almost perfectly
symmetric  with respect to $q=\pi/2$. This doubling of the unit cell results
from the softening at $q=4\kF$, and is a typical signature of the CDW
state. Except for a smaller velocity $v_\sigma$, the spin structure factor in
Fig.~\ref{fig:dynamics-vs-xi-V9} is similar to the metallic regime (\ie,
gapless), see for example Fig.~\ref{fig:dynamics-vs-xi-V6}(c).

\subsection{Single-particle spectral function}

\begin{figure}
   \includegraphics[width=0.45\textwidth]{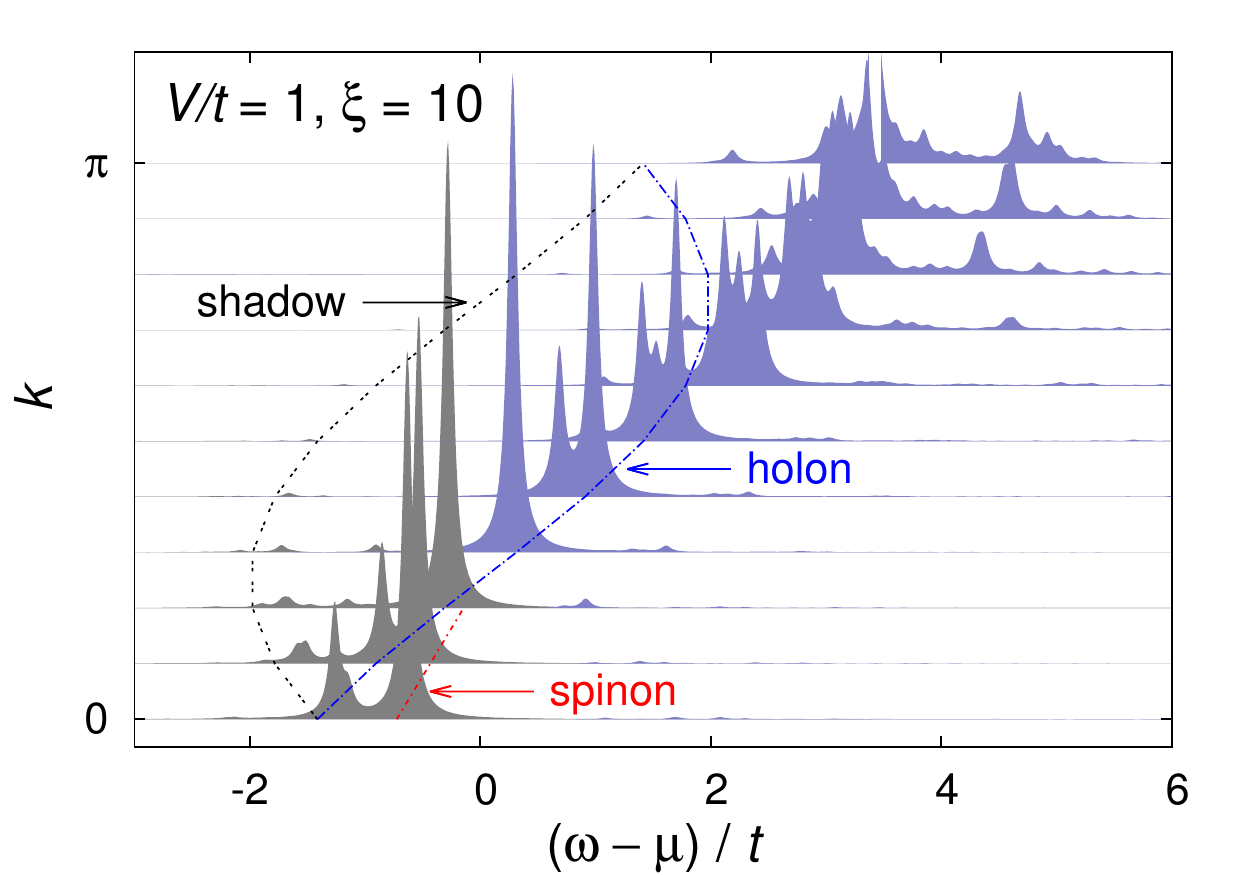}
   \caption{\label{fig:akw-vs-xi-V1} 
     (Color online)
     Single-particle spectral function  $A(k,\om-\mu)$ for $n=0.5$, $V/t=1$     
     and $\xi=10$ from exact diagonalization with $L=20$. We used an artificial
     broadening of $0.05t$. The curves marked spinon, holon
     and shadow are explained in the text.
   }
\end{figure}

The single-particle spectrum is of particular interest in the search for
experimental realizations of LLs because it can reveal the signatures of
spin-charge separation (spinon and holon
bands).\cite{PhysRevB.47.6740,PhysRevB.46.15753} Although LL theory is a
low-energy description, spin-charge separation may be observed up to rather
high energies. For example, spinon and holon bands are visible over an energy
range of the order of the bandwidth in the Hubbard
model,\cite{PhysRevLett.77.1390,PhysRevLett.92.256401,PhysRevB.73.165119} and
also experimentally for TTF-TCNQ\cite{PhysRevLett.88.096402} and 1D
cuprates.\cite{Ki.Ko.Ro.Oh.06,PhysRevLett.77.4054} In contrast, such
clear features of spin-charge separation seem to be absent in recent
measurements on self-organized gold chains, although the density of states
reveals the scaling expected for a LL.\cite{AuchainsNature,PhysRevB.83.121411}

To understand the role of the interaction range and small values of $K_\rho$,
we calculate the single-particle spectral function $A(k,\om-\mu)$
[Eq.~(\ref{eq:akw})] for different values of $V$ and $\xi$. To simplify the
interpretation of the complex structures, we use exact diagonalization on
chains with $L=20$ sites, and use a different graphical representation.

\begin{figure}
   \includegraphics[width=0.45\textwidth]{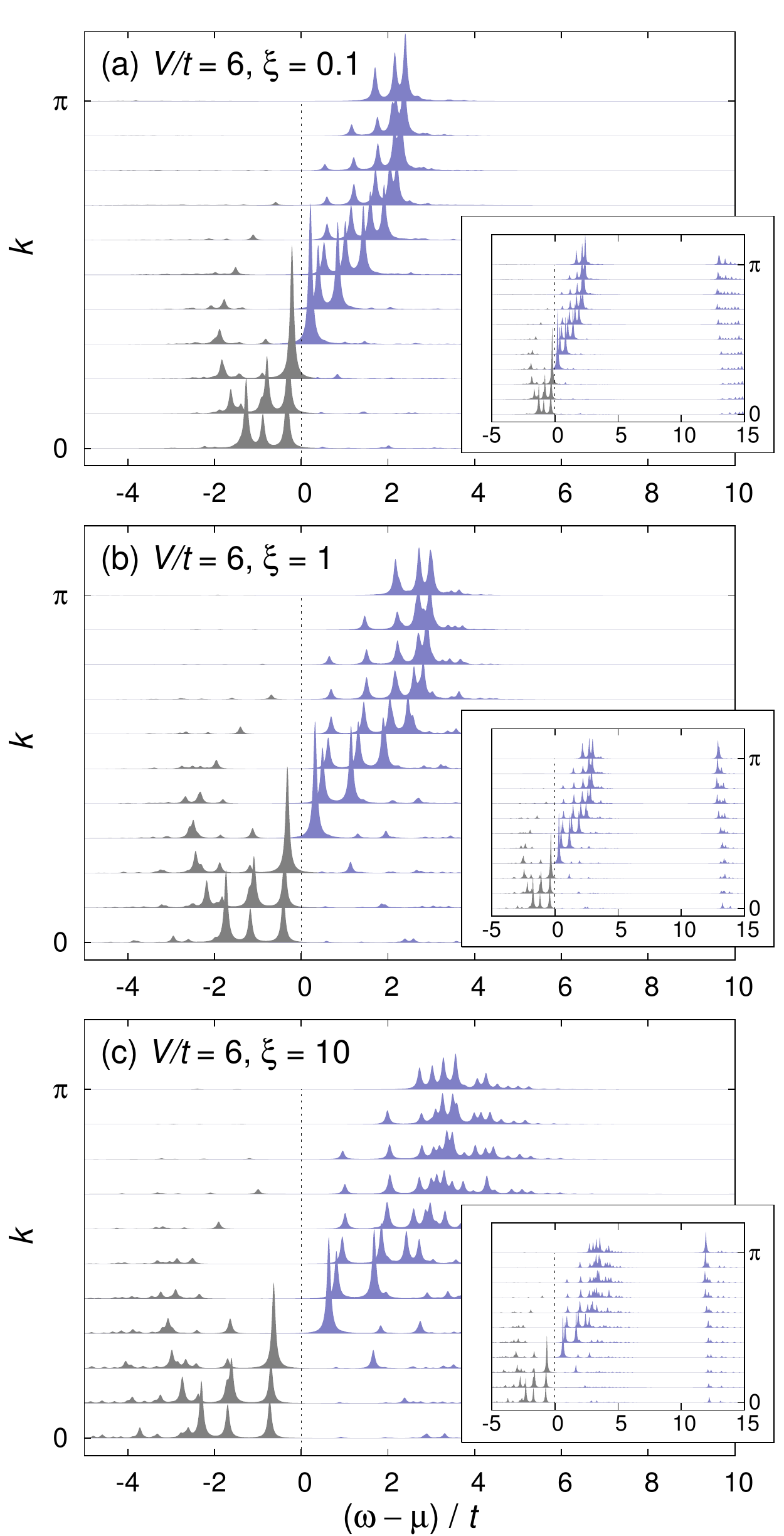}
   \caption{\label{fig:akw-vs-xi-V6} 
     (Color online)
     Single-particle spectral function  $A(k,\om-\mu)$ for $n=0.5$, $V/t=6$     
     and different screening lengths $\xi$ from exact diagonalization with $L=20$.
     Insets: full energy range, revealing the upper Hubbard band.
   }
\end{figure}

Figure~\ref{fig:akw-vs-xi-V1} shows the single-particle spectrum in the
Hubbard regime for $V/t=1$ and $\xi=10$. To highlight the spinon, holon and
shadow bands previously observed for the Hubbard model away from half
filling,\cite{PhysRevLett.77.1390,PhysRevLett.92.256401,PhysRevB.73.165119}
we include the holon and shadow band dispersions for the $U=\infty$ Hubbard
model,\cite{PhysRevLett.92.256401} $-2t \cos(|k| + \kF)$ and
$-2t\cos(|k| - \kF)$, as well as a linear spinon branch $v_\sigma (k - \kF)$
with $v_\sigma$ determined from $S_\sigma(q,\omega)$.  These analytical
results have well-defined corresponding excitations in the numerical spectra,
and establish the signatures of spin-charge separation in the Hubbard regime
of the phase diagram.  The spectral weight of the shadow band at large $k$ is
rather small in Fig.~\ref{fig:akw-vs-xi-V1}. The finite spectral weight
between the holon and spinon excitation peaks is due to the finite system
size.\cite{PhysRevLett.92.256401}

\begin{figure}
   \includegraphics[width=0.45\textwidth]{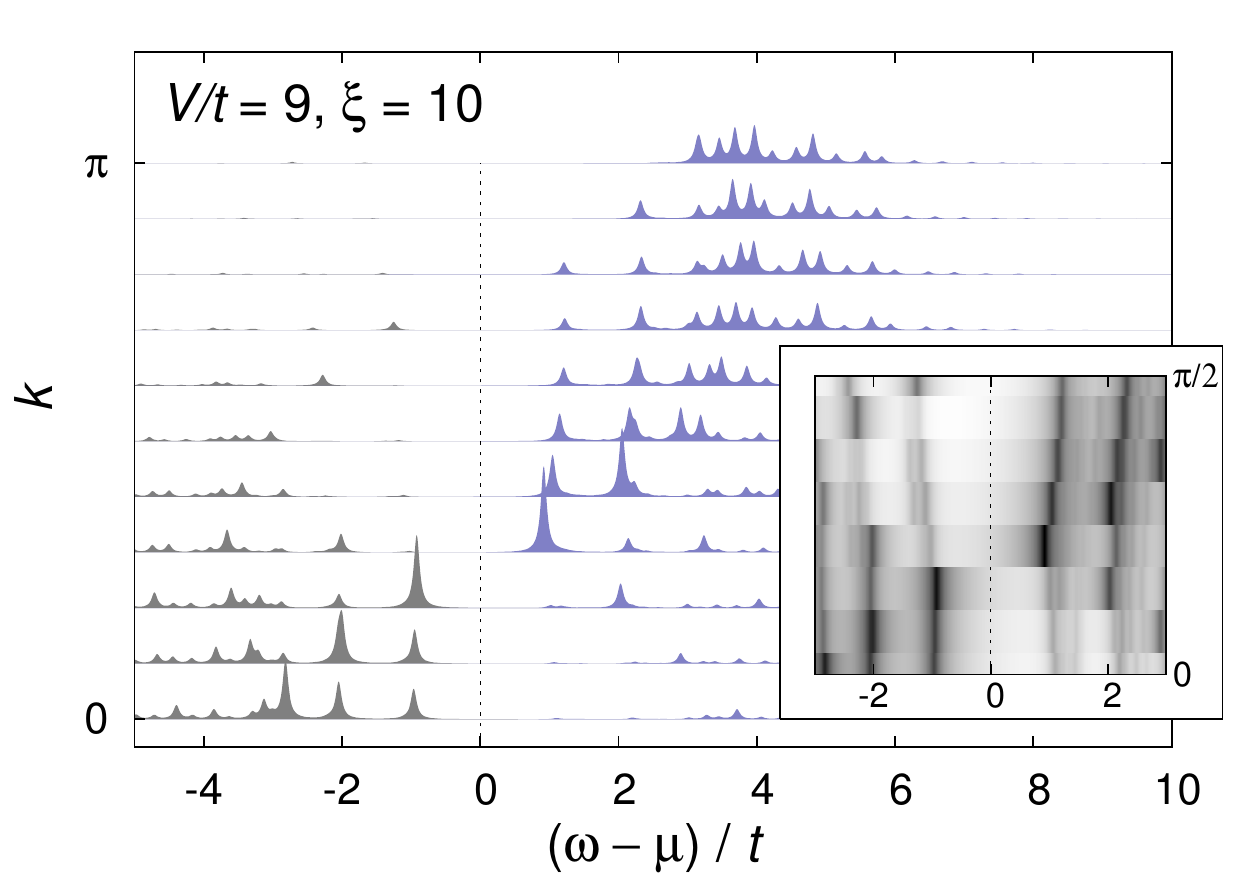}
   \caption{\label{fig:akw-vs-xi-V9} 
     (Color online)
     As in Fig.~\ref{fig:akw-vs-xi-V1} but for $V/t=9$ and $\xi=10$,
     corresponding to the insulating CDW phase, see Fig.~\ref{fig:krho}.
     The inset shows a logarithmic density plot of the spectrum, revealing
     backfolded shadow bands related to the $4\kF$ charge order.
   }
\end{figure}

Taking $V/t=6$, we can study the spectral function across the Hubbard,
non-Hubbard and dominant $4\kF$ regimes with increasing $\xi$.  The results
are shown in Fig.~\ref{fig:akw-vs-xi-V6}, and reveal that the signatures of
spin-charge separation are fully preserved.  Whereas the holon dispersion
reflects the noticeable increase of the charge velocity with increasing
$\xi$, see Fig.~\ref{fig:dynamics-vs-xi-V6}, the spinon excitations remain
virtually unchanged by the interaction range, again in accordance with the
results for $S_\sigma(q,\om)$ in Fig.~\ref{fig:dynamics-vs-xi-V6}. The
spectral weight of the shadow band is significantly enhanced compared to
$V/t=1$ (Fig.~\ref{fig:akw-vs-xi-V1}). On approaching the strong-coupling
region at larger $\xi$, the upper Hubbard band (visible in the insets of
Fig.~\ref{fig:akw-vs-xi-V6}) becomes almost completely flat. Similar to
Fig.~\ref{fig:akw-vs-xi-V6}, a gap is visible at $\kF$ in
Fig.~\ref{fig:akw-vs-xi-V6}(c) [and also in (b) but much smaller]; we have
verified that this gap is a finite-size effect.

Our findings in the metallic region of the phase diagram are consistent with
the experimentally observed coexistence of a small $K_\rho$ (implying
extended-range interactions) with signatures of spin-charge separation in
photoemission measurements; a good example is
TTF-TCNQ.\cite{PhysRevLett.88.096402} On the other hand, the finite
interaction range does not provide an explanation for the possible absence
of clear spin-charge separation in self-organized gold chains.\cite{AuchainsNature,PhysRevB.83.121411} We comment on
the latter case in the conclusions.

Finally, we show in Fig.~\ref{fig:akw-vs-xi-V9} the single-particle spectrum
in the insulating CDW phase at $V/t=9$ and $\xi=10$. The dynamical charge and
spin structure factors for these parameters were presented in
Fig.~\ref{fig:dynamics-vs-xi-V9}. We find a charge gap (equal to
$0.2(1)$ in the thermodynamic limit), and backfolded shadow bands related to
the $4\kF$ charge order which are visible in the inset of
Fig.~\ref{fig:akw-vs-xi-V9}. The spectrum appears to evolve continuously
across the CDW transition. In particular, the holon band is well visible in
Fig.~\ref{fig:akw-vs-xi-V9}, whereas it has been found to separate into two
domain walls for much stronger Coulomb interaction.\cite{PhysRevB.75.125116}
The single-particle spectrum of a quarter-filled CDW state has also been
calculated using the bosonization method.\cite{essler.tsvelik.2002} In the
absence of dimerization, no singularities exist near $\kF$ (note that in our
numerical calculations, we cannot distinguish between singularities and
excitation peaks of finite width). The spectrum may also depend on the
details of the interaction potential.

\section{Conclusions}\label{sec:conclusions}

In this work, we have studied the effects of the electron-electron
interaction range in one dimension using exact numerical methods. We have
obtained the Luttinger liquid interaction parameter $K_\rho$ as a function of
the Coulomb matrix element $V$ and the screening length $\xi$ which, in
combination with Luttinger liquid theory, defines the phase diagram of the
model. In addition to the Hubbard regime $1\geq K_\rho \geq 1/2$, we have
explored the non-Hubbard regime $K_\rho<1/2$, the case $K_\rho<1/3$ with
dominant $4\kF$ charge correlations, and the insulating CDW state which
exists at quarter filling for $K_\rho<1/4$. We identified an important length
scale $1/\kF$ for $K_\rho$; $K_\rho$ strongly depends on the screening length
for $\xi\lesssim 1/\kF$, whereas it decays very slowly for $\xi\gg1/\kF$. Our
results indicate that the lattice model with a finite (but possibly large)
interaction range can be described by Luttinger liquid theory if higher-order
umklapp terms are taken into account. This case is therefore distinct from
the unscreened $1/r$ potential which falls outside the Luttinger liquid
description.\cite{PhysRevB.45.8454,PhysRevLett.71.1864} For the unscreened
potential, numerical results suggest the existence of a metallic quasi Wigner
crystal state with $K_\rho=0$.\cite{PhysRevB.60.15654} For our choice of a
screened Coulomb potential, which is both convex and monotonically decreasing
with increasing distance, $K_\rho$ always decreases with increasing
interaction strength or range, as compared to enhanced metallic behavior
observed in extended Hubbard models as a result of competing nearest-neighbor
and next-nearest neighbor interactions.  Interestingly, the small values of
$K_\rho\approx1/4$ observed in recent experiments on gold chains, as well as
previously in quantum wires, carbon nanotubes and quasi-1D materials, can
only be achieved for large values of the interaction strength and/or range.

We have calculated the static and dynamical charge and spin correlation
functions, and found good agreement with the expectations based on Luttinger
liquid theory.  Upon decreasing $K_\rho$ by increasing $V$ and/or $\xi$,
$4\kF$ charge correlations become strongly enhanced, reminiscent of although
not identical to the quasi Wigner crystal. Our results for the real-space
density-density correlations are consistent with Luttinger liquid behavior on
length scales beyond the screening length and deviations on smaller length
scales.

The $4\kF$ correlations lead to a pronounced Bragg peak in the dynamical
density structure factor. The interaction range strongly modifies the
velocity of long-wavelength charge excitations, whereas the spin velocity
only depends on the onsite repulsion.  Throughout the Luttinger liquid phase,
spin-charge separation is clearly visible in the single-particle
spectrum. Finally, in the insulating charge-density-wave phase, we observe
backfolded shadow bands.

An important question to be addressed in future work, motivated by
experiments on self-organized gold chains,\cite{AuchainsNature,PhysRevB.83.121411}
is the impact of spin incoherence on the spinon and holon signatures in
photoemission spectra. The
energy scales for low-energy charge and spin excitations are determined by
the corresponding velocities $v_\rho$ and $v_\sigma$. As explicitly shown in
this work, $v_\rho$ increases with increasing $\xi$, whereas $v_\sigma$ does
not depend on the interaction range. Therefore, the charge and spin energy
scales can be well separated for sufficiently large $\xi$. In the regime
$v_\rho\gg v_\sigma$, the $2\kF$ spin correlations can be suppressed at
finite temperatures, whereas the charge sector remains
coherent.\cite{RevModPhys.79.801} This scenario may explain the rather
incoherent angle-resolved spectrum of gold chains, which at the same time
show clean LL power-law behavior in the angle-integrated density of
states.\cite{AuchainsNature}

{\begin{acknowledgments}%
    We thank D. Baeriswyl, R. Claessen, S. Eggert, S. Ejima, 
    F. Essler, H. Fehske, V. Meden, J. Sch\"afer, and D. Schuricht for helpful discussions. This work
    was supported by the DFG Grants No.~FOR1162 and WE 3639/2-1, as well as
    by the Emmy Noether Programme.  We are grateful to the LRZ Munich and the
    J\"ulich Supercomputing Centre for generous computer time.
\end{acknowledgments}}

\vspace*{2em}
\section*{\MakeUppercase{Appendix: CTQMC}}
\appendix*

The general formulation of the weak-coupling CTQMC method allows the
simulation of problems with long-range interactions in imaginary time and/or
space.\cite{Ru.Sa.Li.05,assaad:035116,Gull_rev} Retarded interactions (\ie,
nonlocal in time), which essentially correspond to the electron-phonon
problem, have been considered in
Refs.~\onlinecite{assaad:155124,Marcin10,Hohenadler10a}.  In this appendix,
we provide technical details for the application of the CTQMC method to a
Hamiltonian of the form~(\ref{eq:ham}).

Although such simulations are in principle straight-forward, we have
encountered difficulties which are ultimately related to the strong-coupling
character of the problem considered in this paper. The algorithm is quite
similar to the case of electron-phonon interactions, and has been implemented
both at finite temperatures and at $T=0$ (with a projection parameter
$\theta$).\cite{assaad:035116}

Our starting point is Eq.~(\ref{eq:ham}),  which we write as
\begin{equation}
  H = \sum_{k} \epsilon(k) \on_{k} +
  \overline{V} \sum_{ir}   P(r) \left( \on_i  - n
  \right)  \left( \on_{i+r}  - n  \right) .
\end{equation}  
Here $n$ is the average density, the interaction accounts for fluctuations
around the paramagnetic saddle point, and $P(r)$ is a probability
distribution; we also defined $\overline{V}=\sum_r V(r)$.  During the
simulation, vertices corresponding to interactions over a distance $r$ are
proposed with probability $P(r)=V(r)/\overline{V}$.

To circumvent the negative sign problem, and following
Ref.~\onlinecite{assaad:035116},  we rewrite the interaction as
\begin{equation}\label{eq:newV}
  \oh\overline{V} \sum_{ir\sigma\sigma's}   P(r)
  \left( \on_{i\sigma}  - \frac{n}{2}  + s \delta  \right)  \left(
    \on_{i+r\sigma'}  - \frac{n}{2}  - s \delta\right) .
\end{equation}
Here we have introduced an Ising variable  $s = \pm 1$. Up to a constant,
Eq.~(\ref{eq:newV}) is equivalent to the original interaction. To avoid the
sign problem for $\overline{V}>0$ we have the condition 
$n/2  +  \delta  > 1$.

The average expansion order, which determines the computer time, can be
evaluated within the finite temperature approach, giving
\begin{equation}
  \langle M \rangle =  \beta \overline{V} L \bigg[ 4 \delta^2 -  \sum_{r } P(r )
    \langle \left( \on_{0}  - n  \right)  \left( \on_{r}  - n  \right)  \rangle   \bigg]\,.
\end{equation}
The fact that the form~(\ref{eq:newV}) is beneficial for the simulations at
quarter filling and for rather strong interactions confirms an empirically
derived rule. In order to obtain optimal results away from half filling, it
is often useful to increase the value of $\delta $ at the expense of a larger
average expansion order. With the above formulation, we were able to extend
the parameter regime of applicability for the weak-coupling CTQMC method, and
exemplary results are shown in Fig.~\ref{fig:dynamics-vs-xi-V1}. However, the
strong-coupling regime remains out of reach.


\end{document}